\documentclass[preprint,12pt]{elsarticle}

\usepackage{amssymb}

\usepackage{lineno}
\usepackage{graphicx}
\usepackage{subfigure}
\usepackage{subfig}
\journal{JINST}

\biboptions{sort&compress}

\begin{document}

\begin{frontmatter}

\title{A study of liquid argon detector's $n$/$\gamma$ discrimination capability with PMT or SiPM readout}

\renewcommand{\thefootnote}{\fnsymbol{footnote}}
\author{
L.~Wang$^{a,b}$, 
Y.~Lei$^{c}$,
M.Y~Guan$^{a,b,d}$,
T.A.~Wang$^{e}$, 
C.~Guo$^{a,b,d}\footnote{Corresponding author. Tel:~+86-1088236256. E-mail address: guocong@ihep.ac.cn (C.~Guo). }$, 
J.C.~Liu$^{a,b,d}$, 
C.G.~Yang$^{a,b,d}$, 
X.H.~Liang$^{f}$, 
Y.D.~Chen$^{c}\footnote{Corresponding author. Tel:~+86-13661260225. E-mail address: chenyuede@fjut.edu.cn (Y.D.~Chen).}$
}

\address{
	${^a}${Experimental Physics Division, Institute of High Energy Physics, Chinese Academy of Sciences, Beijing, China}
	
	${^b}${School of Physics, University of Chinese Academy of Sciences, Beijing, China}
	
	${^c}${State Key Laboratory of High Power Semiconductor Laser, College of Physics, Changchun University of Science and Technology, Changchun, Jilin, China}
	
    ${^d}${School of Electronic, Electrical Engineering and Physics, Fujian University of Technology, Fuzhou, China}

	${^e}${State Key Laboratory of Particle Detection and Electronics, Beijing, China}
	
	${^f}${Astro-particle Physics Division, Institute of High Energy Physics, Chinese Academy of Science, Beijing, China}
}


\begin{abstract}
	Liquid Argon (LAr) is used as a target material in several current and planned experiments related to dark matter direct searching and neutrino detection. Argon provides excellent Pulse Shape Discrimination (PSD) capability which could separate the electron recoil backgrounds from the expected nuclear recoil signals. This essay simulated the PSD capability of an LAr detector when PMTs or three kinds of SiPMs are used as photosensors based on the experimental data. The results show that the J-60035 SiPM could help the LAr detector achieve the highest PSD capability event though SiPM's After-Pulse (AP) and Cross-Talk (CT) deteriorate its PSD capability. In addition, the results also show that the effect from AP is greater than CT. This is instructive for selecting photosensors for LAr detectors.
\end{abstract}

\begin{keyword}
	Liquid argon, SiPM, PMT, F90, PSD 
\end{keyword}

\end{frontmatter}

\section{Introduction}\label{sec:section1}

Liquid argon (LAr) has been widely used in dark matter direct searching experiments~\cite{DarkSide-20k, Deap2008} and neutrino detecting experiments~\cite{Gerda, DUNE1}. The excellent pulse shape discrimination (PSD) capability of LAr detectors can significantly lower the background event rate caused by overwhelming electron recoils~\cite{PSDpaper1, PSDpaper2, PSDpaper3}. Silicon Photon Multiplier (SiPM), as the next-generation photosensors for scintillator detectors, have many advantages over traditional Photon Multiplier Tubes (PMTs), including competitive prices, lower operating voltages, lower radio-active background, and higher quantum efficiency~\cite{SiPM}. Thus, SiPM-based photosensors have been expected to play a significant role in further low-threshold experiments\cite{DarkSide-20k, nEXO}.

Previous works~\cite{PreviousWork1, PreviousWork2, PreviousWork3} have proved that SiPM is a good photosensor candidate for LAr detectors. However, the difference in the LAr detector's PSD capability when using SiPMs or PMTs as photosensors is barely studied, but theoretically exists. Compared with PMTs, the wide Single PhotoElectron (SPE) waveform and the higher probability of correlated signals of SiPM may reduce the PSD capability, while the much better SPE energy resolution of SiPM may improve its PSD capability. In this work, the authors will discuss how a photosensor’s SPE energy resolution, pulse shape, and correlated signals affect the LAr detector’s PSD capability. 

This study was conducted under the hypothesis that the same number of photoelectrons are detected in the case of PMT or SiPM readout and the effects of light propagation have not been taken into consideration, only the intrinsic effects due to argon scintillation and sensor response are considered.

\section{Prompt fraction method}\label{sec:section2}

The prompt fraction method is commonly used for PSD in the LAr detector. The internal mechanism of the prompt fraction method is that there are two scintillation processes to emit photons with different decay times. They are the singlet state ($^1 \Sigma_u^+$) with a decay time of $\sim$7~ns and the triplet state ($^3 \Sigma_u^+$) with a decay time of $\sim$1.6 $\mu$s separately ~\cite{LAr_Luminescence}. Heavy particles like neutrons, which tend to produce Nuclear Recoil (NR) signals, can proportionately scintillate photons less from the triple state when compared to light particles like $\gamma$, which tend to produce Electron Recoil (ER) signals. As a consequence, the pulse shapes are significantly different when different incident particles deposit the same energy in an LAr detector. Thus, the percentage of energy in the first several dozen nanoseconds of a pulse could be calculated to distinguish nuclear recoils from overwhelming electron recoils. The prompt fraction is calculated for
each pulse according to Eq.~\ref{Eq.Fp}: 
\begin{equation}
	f_p=\frac{\int_{T_s}^{T_i}V(t)dt}{\int_{T_s}^{T_e}V(t)dt}
	\label{Eq.Fp}
\end{equation}
Where f$_p$ is the prompt fraction, V(t) is the amplitude of the output pulse in the unit of mV. T${_s}$ and T${_e}$ are the start and end times of a pulse in the unit of ns. T$_i$ is the end time of the prompt fraction of a pulse in the unit of ns. T${_i}$ = 90~ns has been proven to be most effective for n/$\gamma$ discrimination for an LAr detector with PMT readout. With this method, DEAP-3600 achieved 7.6×$10^{-7}$ Electron Recoil Contamination (ERC) for 50\% acceptance at an energy range of 52~keV to 110~keV~\cite{Deap2008}.

\section{SPE performance of a PMT and three SiPMs}\label{sec:section3}

	\begin{table}[htb]
	\centering
	\renewcommand{\arraystretch}{1.5}
	\scriptsize
	\begin{tabular}{cccccc}
		
		\hline
		Photosensor & Bias (V) &  DCR (Hz/inch$^{2}$) & SPE rise time (ns) & CT & AP\\
		
		\hline
		R11065 PMT & 1750&   $\sim$20&	$\sim$10 & negligible & neglible\\
		S14161-6050HS SiPM & 36.3&	$\sim$250&$\sim$90 & $\sim$26\% & $\sim$10\%\\		
		J-60035 SiPM & 28.3&	$\sim$200&	$\sim$230& $\sim$27\% & $\sim$0.3\%\\
		S13370-6050CN SiPM& 48&	$\sim$0.5&	$\sim$35& $\sim$2\% & $\sim$5\%\\
		\hline 
	\end{tabular}
	\caption{\label{table.1} Characteristics of the four photosensors. All parameters were obtained at 87~K. The detail of the Dark Counting Rate (DCR) and the Cross-Talk (CT) and After-Pulse (AP) probabilites measurement can be found in Ref.~\cite{PreviousWorks4,VUV4_SiPM}.
	}
   \end{table}

\begin{figure*}[htbp]
	\centering
	\subfigure{
		\rotatebox{90}{\scriptsize{~~~~~~R11065 PMT}}
		\begin{minipage}[t]{0.299\linewidth}
			\centering
			\includegraphics[width=1\linewidth]{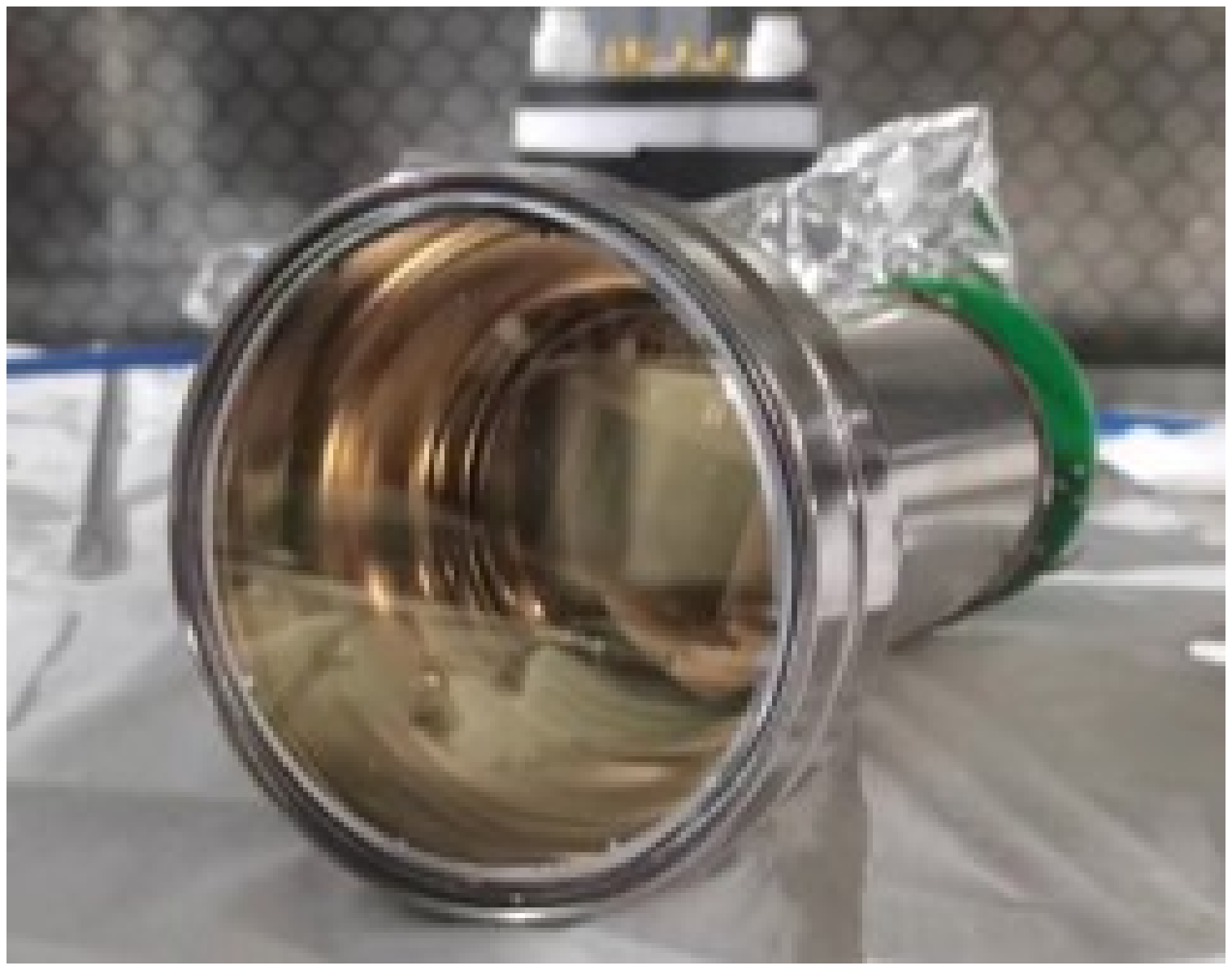}
		\end{minipage}
	}
	\subfigure{
		\begin{minipage}[t]{0.299\linewidth}
			\centering
			\includegraphics[width=1\linewidth]{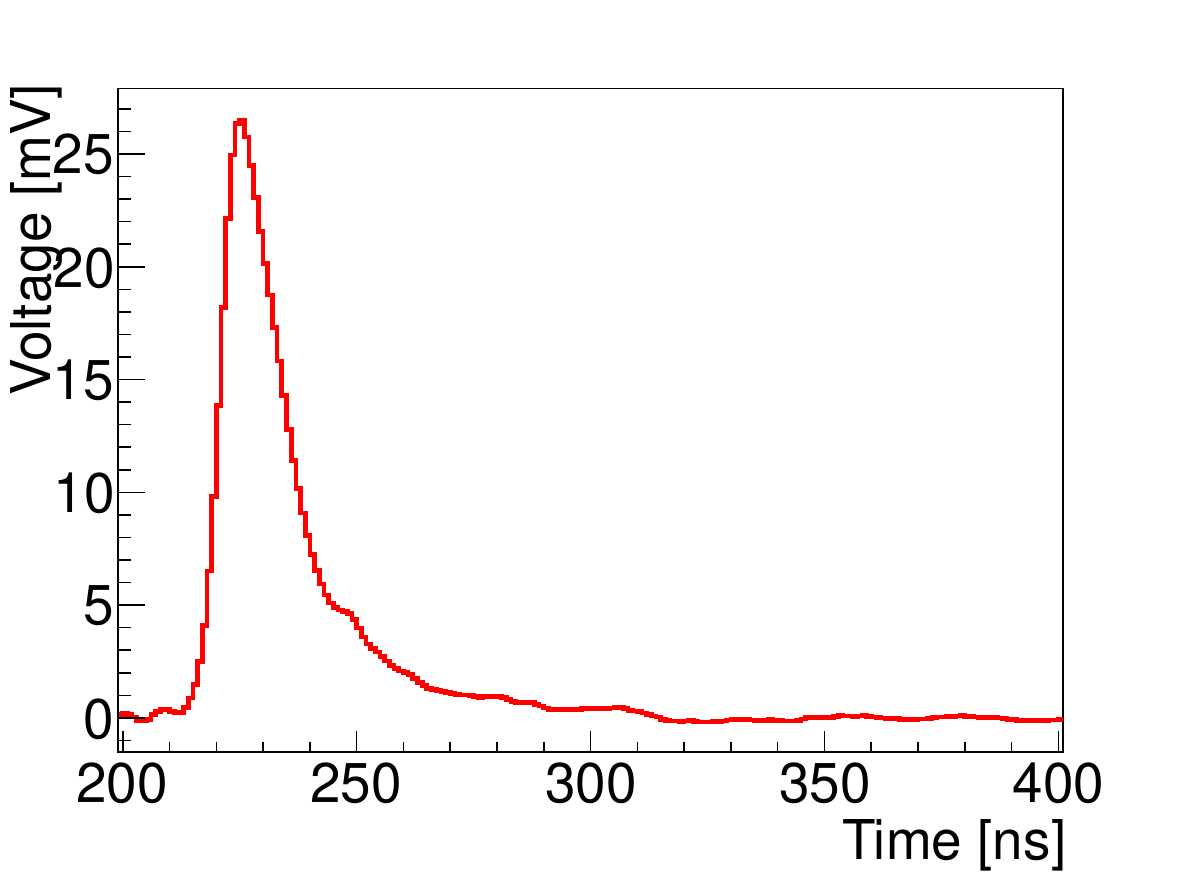}
		\end{minipage}
	}
	\subfigure{
		\begin{minipage}[t]{0.299\linewidth}
			\centering
			\includegraphics[width=1\linewidth]{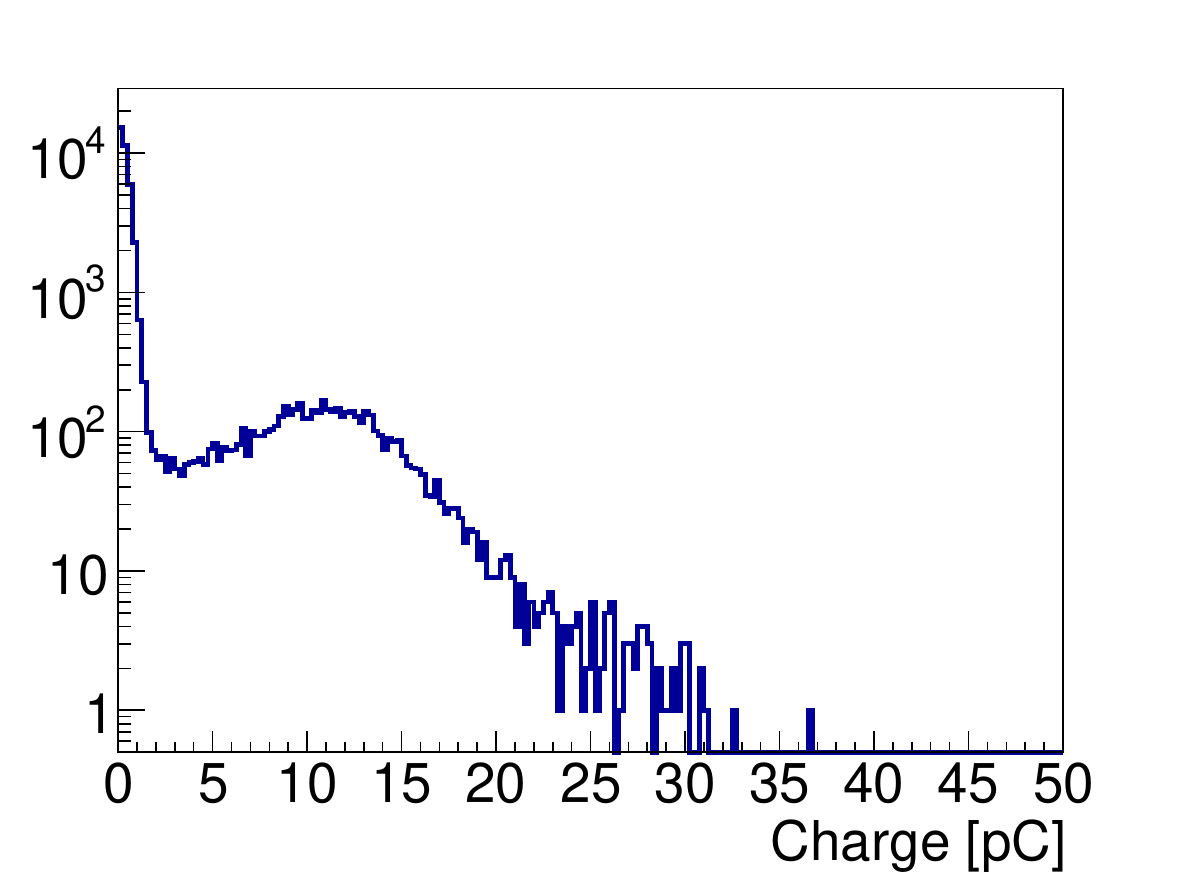}
		\end{minipage}
	}
	
	\subfigure{
		\rotatebox{90}{\scriptsize{S14161-6050HS SiPM}}
		\begin{minipage}[t]{0.299\linewidth}
			\centering
			\includegraphics[width=1\linewidth]{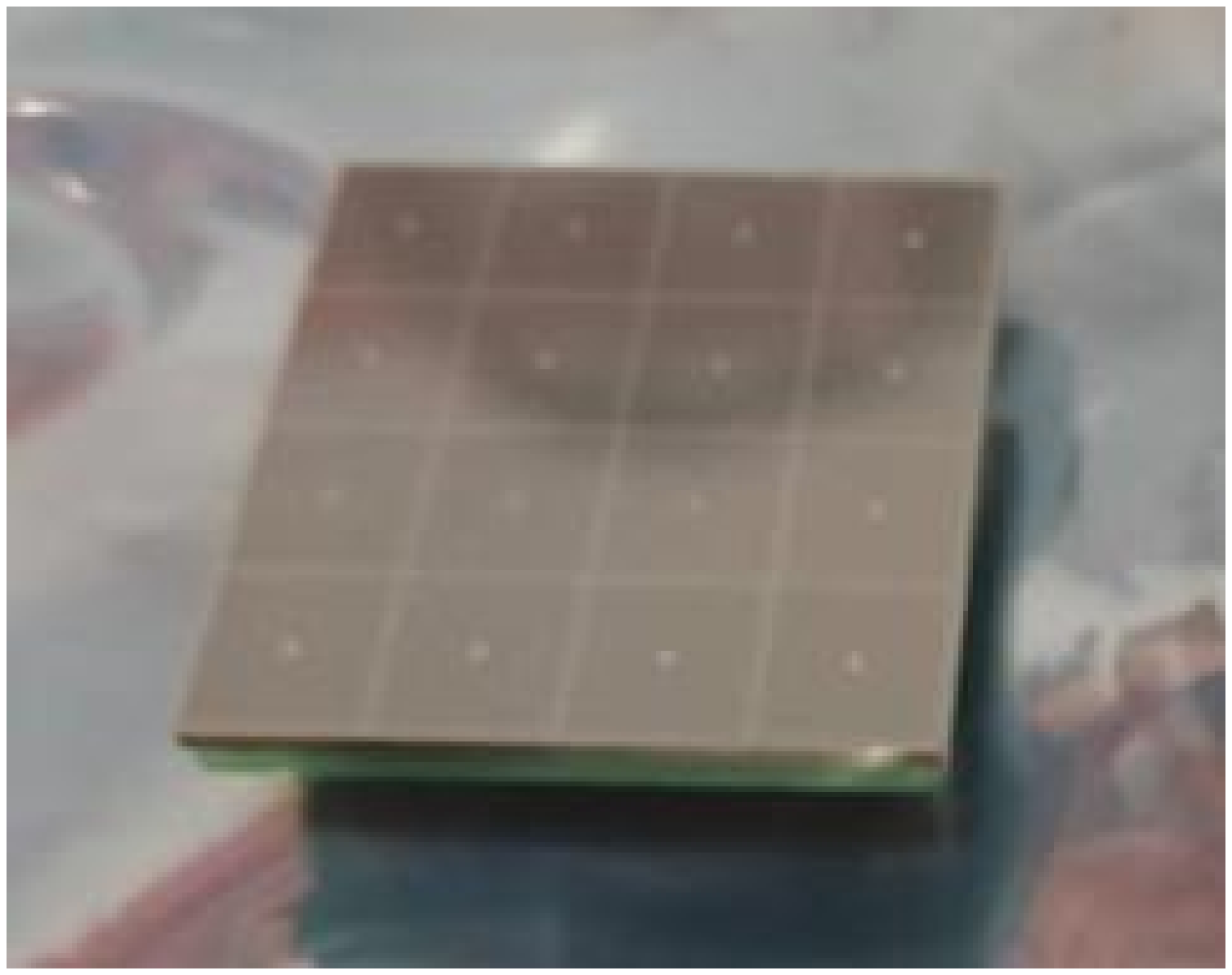}
		\end{minipage}
	}
	\subfigure{
		\begin{minipage}[t]{0.299\linewidth}
			\centering
			\includegraphics[width=1\linewidth]{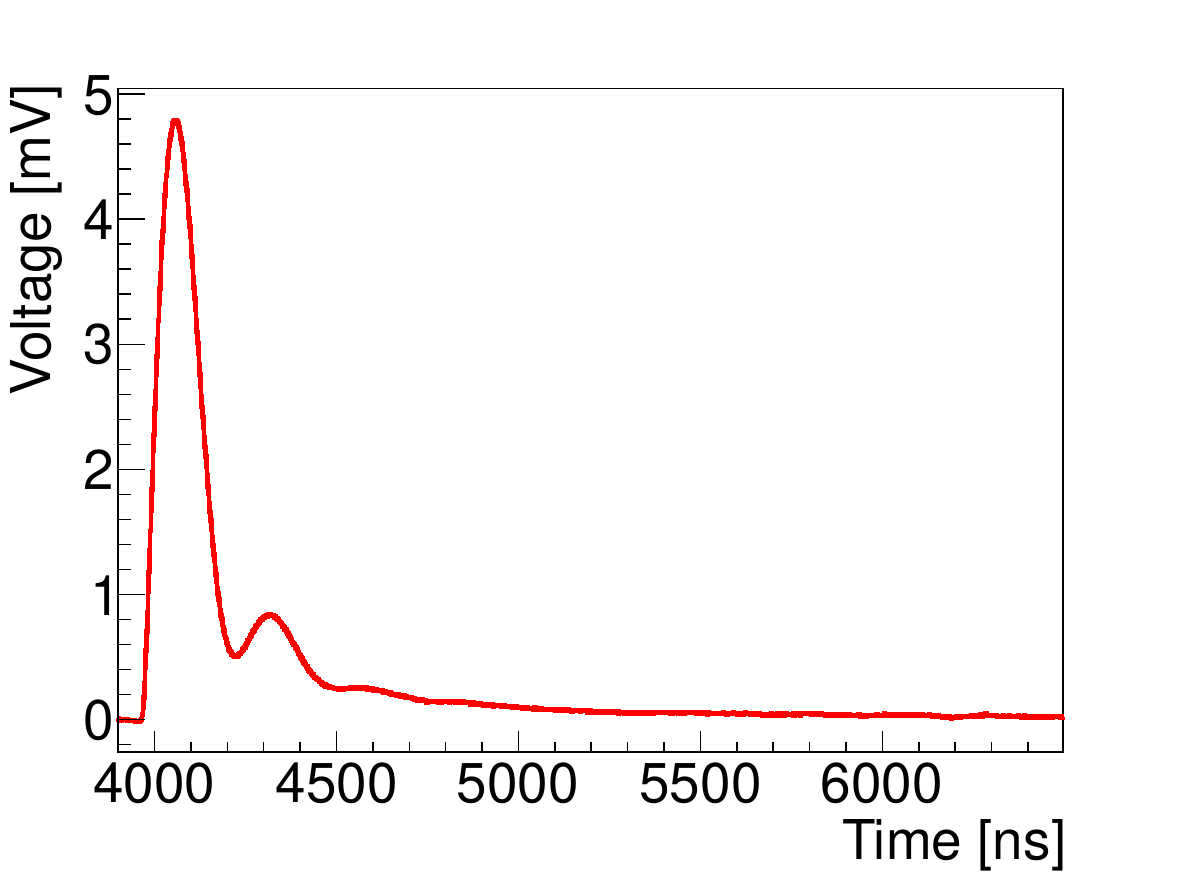}
		\end{minipage}
	}
	\subfigure{
		\begin{minipage}[t]{0.299\linewidth}
			\centering
			\includegraphics[width=1\linewidth]{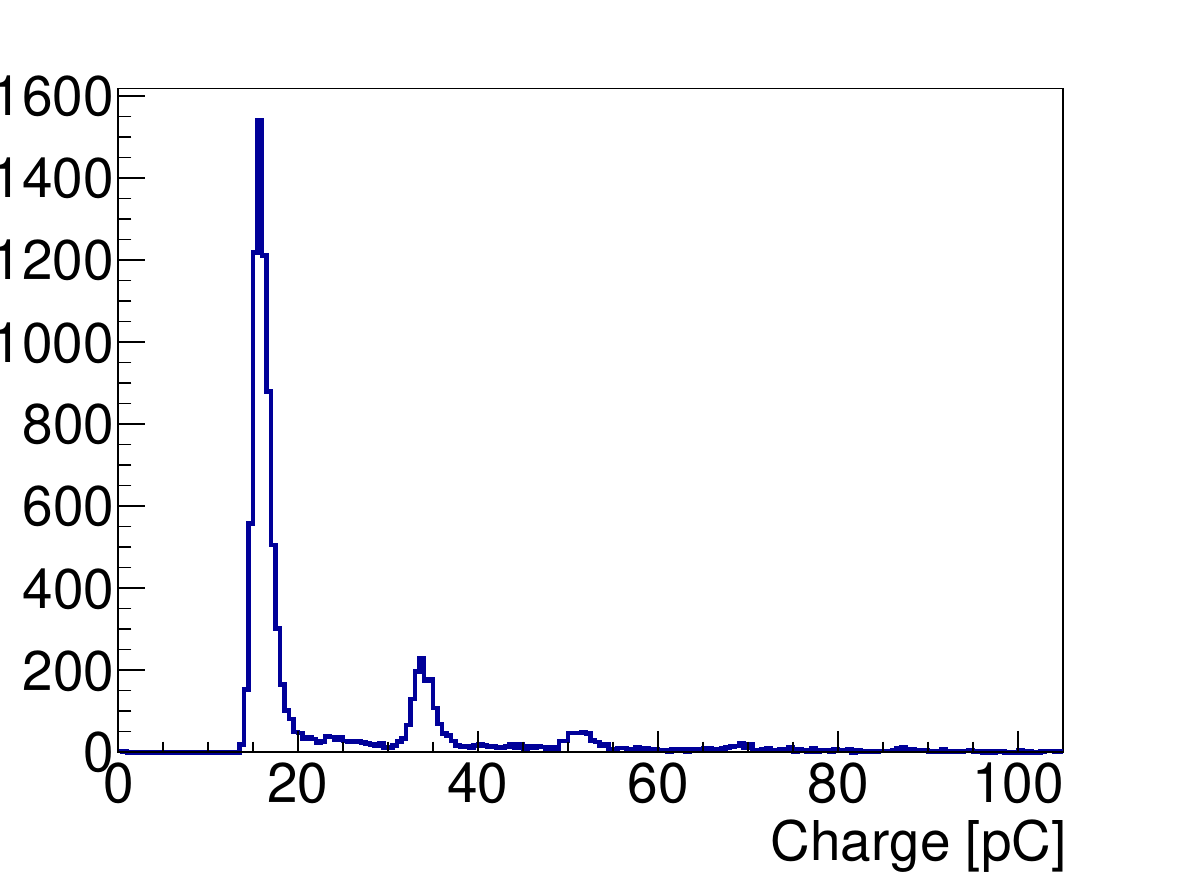}
		\end{minipage}
	}
	
		\subfigure{
		\rotatebox{90}{\scriptsize{~~~~~J-60035 SiPM}}
		\begin{minipage}[t]{0.299\linewidth}
			\centering
			\includegraphics[width=1\linewidth]{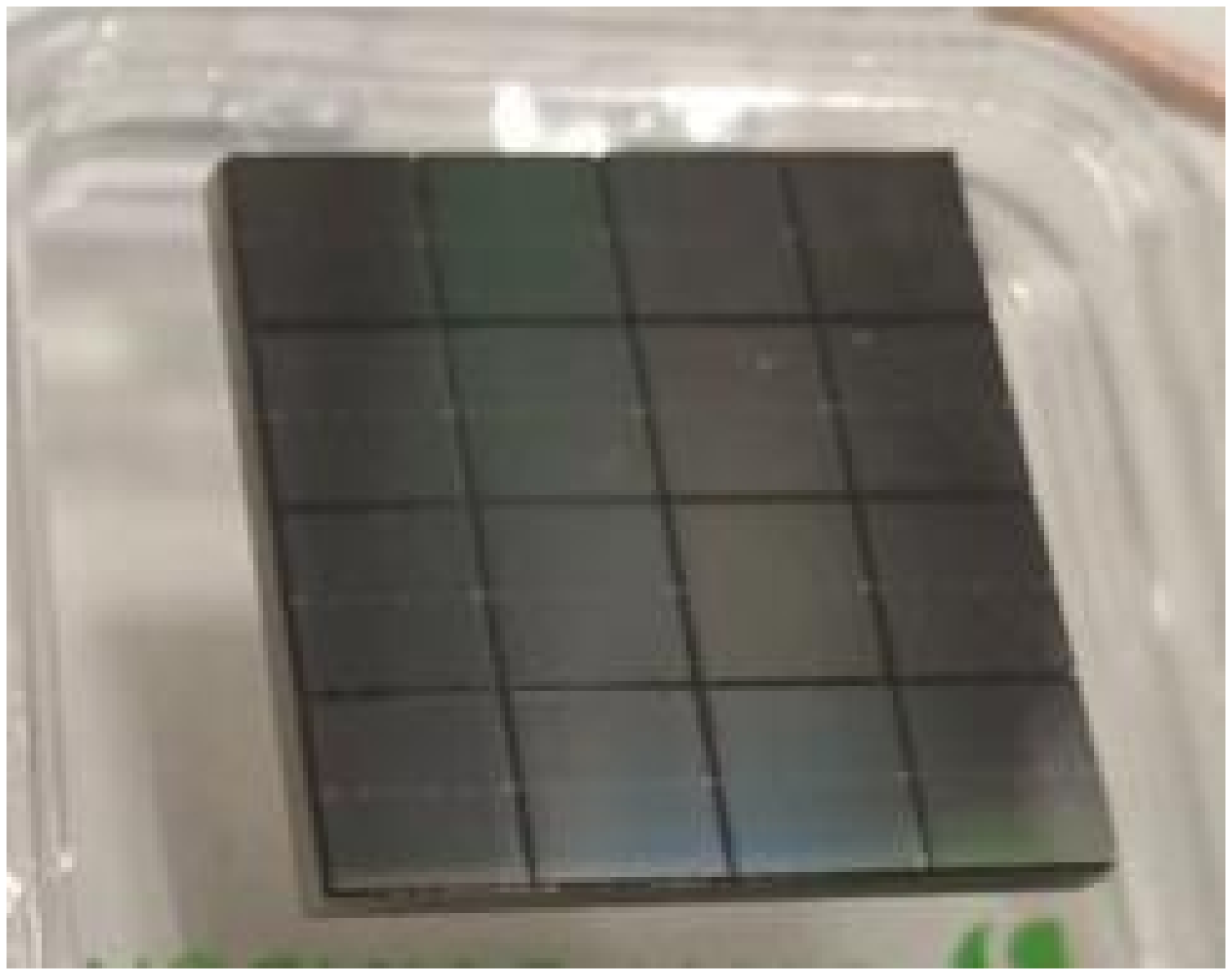}
		\end{minipage}
	}
	\subfigure{
		\begin{minipage}[t]{0.299\linewidth}
			\centering
			\includegraphics[width=1\linewidth]{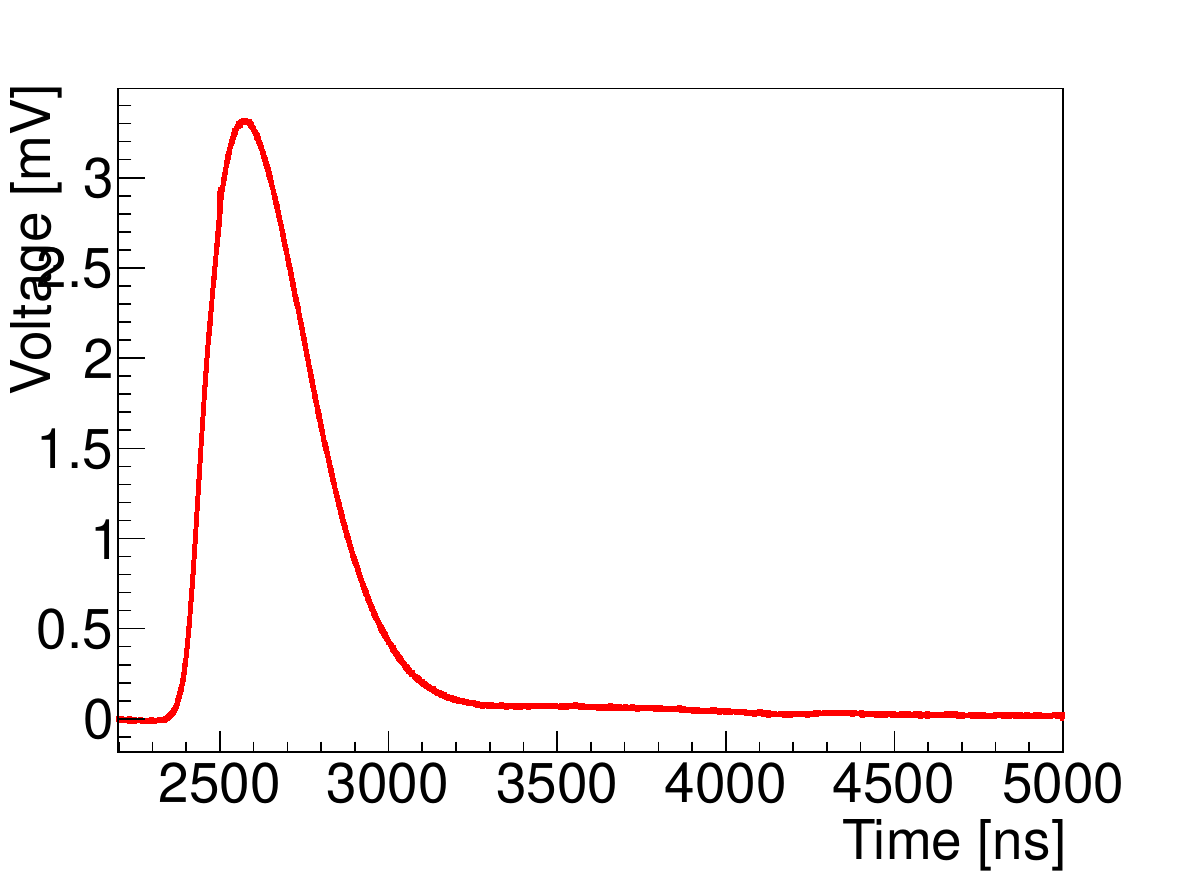}
		\end{minipage}
	}
	\subfigure{
		\begin{minipage}[t]{0.299\linewidth}
			\centering
			\includegraphics[width=1\linewidth]{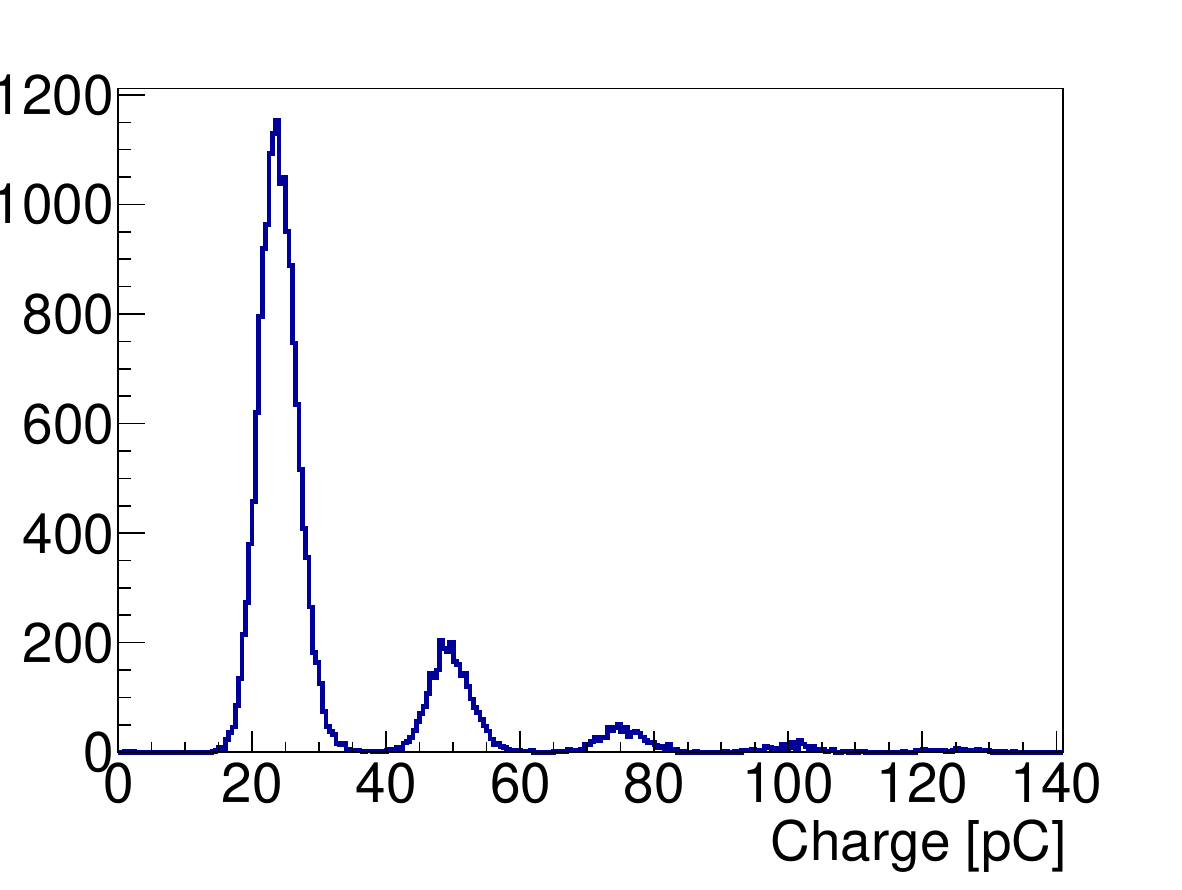}
		\end{minipage}
	}

	\setcounter{subfigure}{0}
		
	\subfigure[Four different photosensors]{
		\rotatebox{90}{\scriptsize{S13370-6050CN SiPM}}
		\begin{minipage}[t]{0.299\linewidth}
			\centering
			\includegraphics[width=1\linewidth]{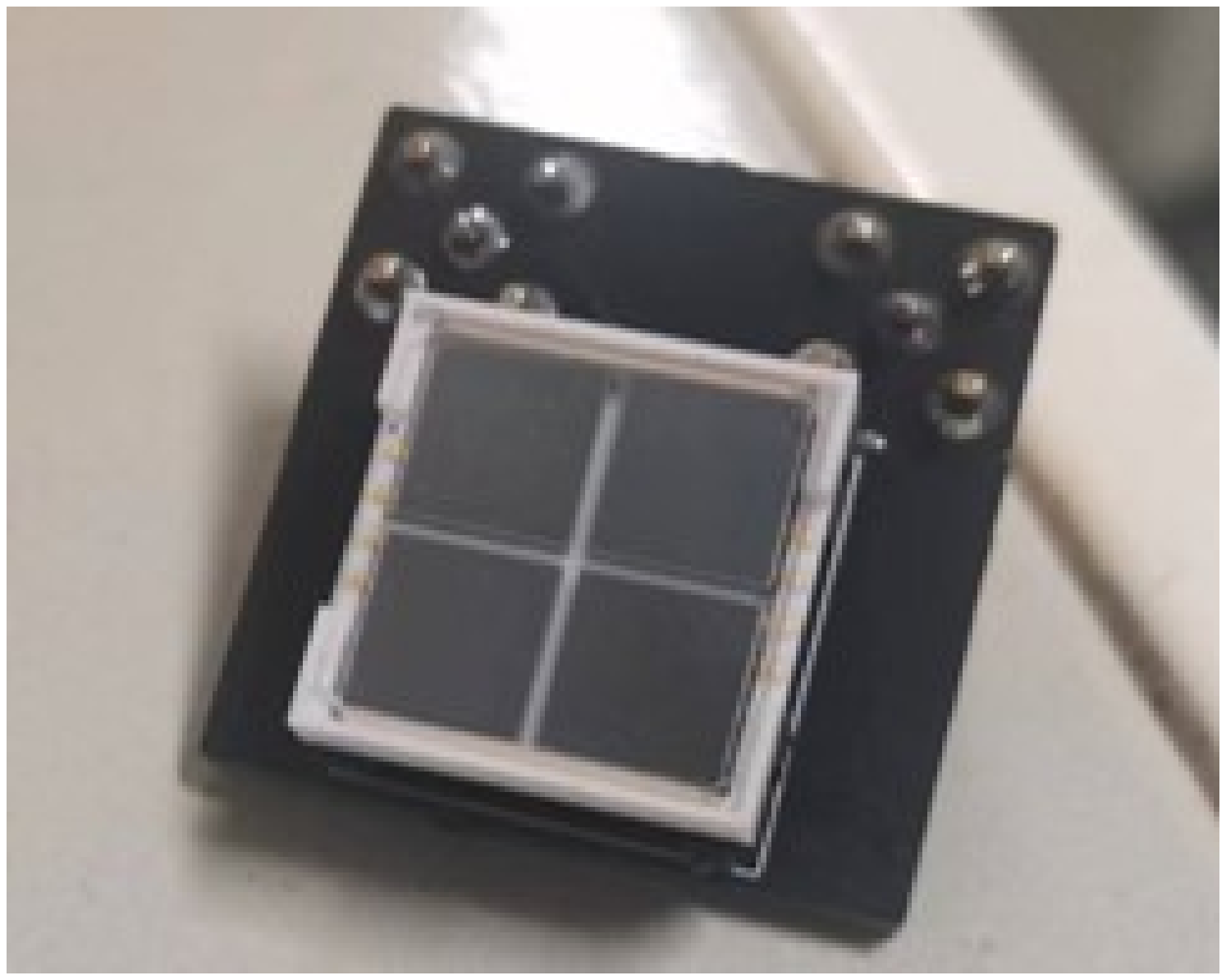}
		\end{minipage}
	}
	\subfigure[Average SPE waveforms respectively]{
		\begin{minipage}[t]{0.299\linewidth}
			\centering
			\includegraphics[width=1\linewidth]{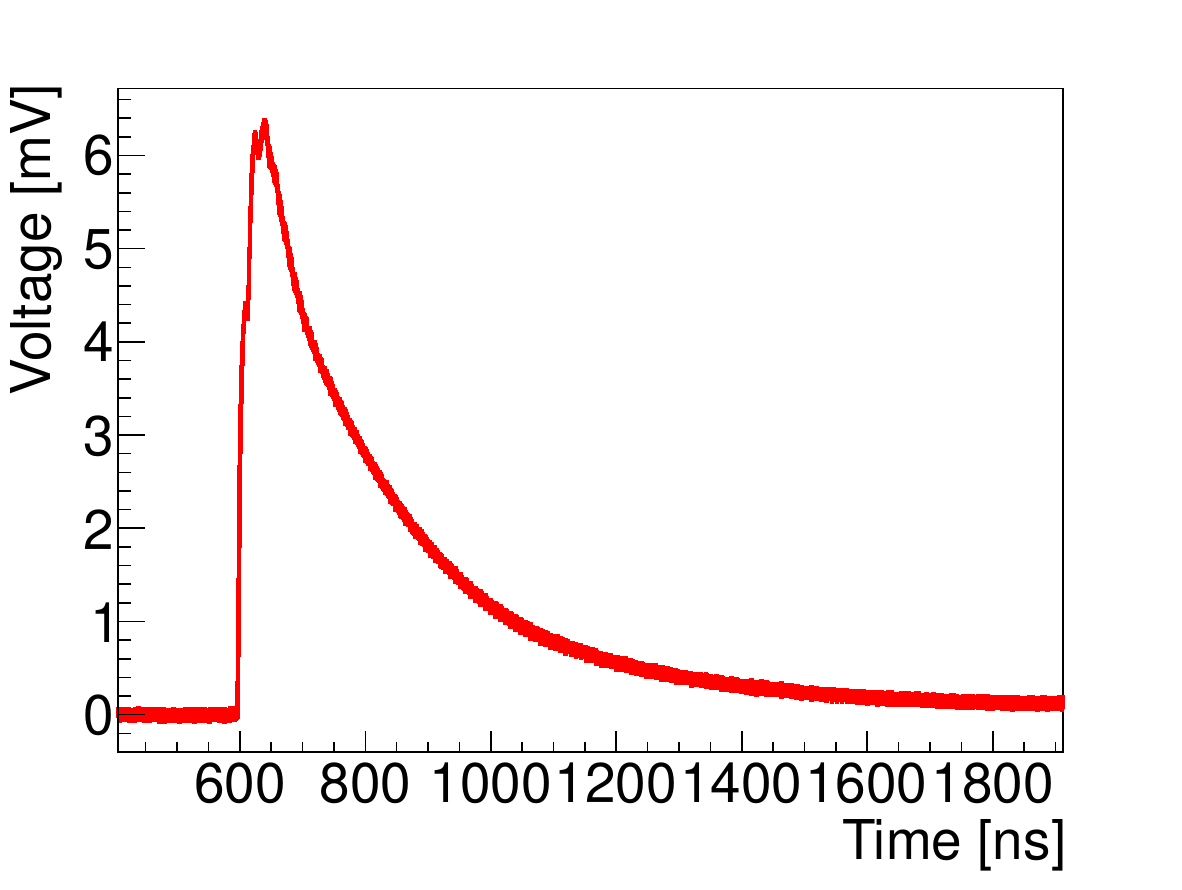}
		\end{minipage}
	}
	\subfigure[Distributions of SPE charge respectively]{
		\begin{minipage}[t]{0.299\linewidth}
			\centering
			\includegraphics[width=1\linewidth]{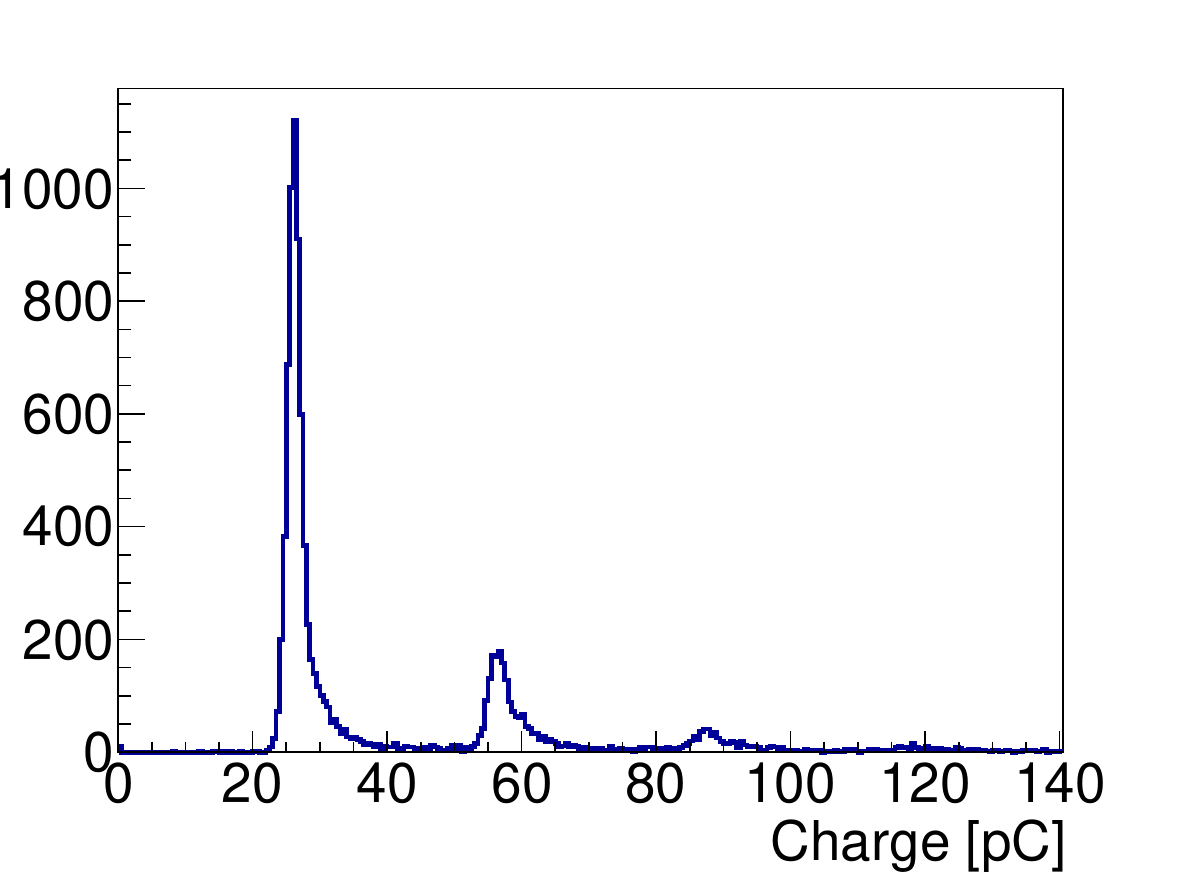}
		\end{minipage}
	}
	
	\caption{Pictures of the four photosensors and their average SPE waveforms and spectra measured in the dark. The second peak in the average SPE waveform of the Hamamatsu S14161-6050HS SiPM near $\sim$4300~ns is due to an electronics mismatch.}
	\label{fig:result_include1}
\end{figure*}

A cryogenic system was built for SiPM SPE measurement at liquid argon temperature and the detail of the experimental setup can be found in Ref~\cite{PreviousWorks4,VUV4_SiPM}. The characteristics of three different types of SiPM, which are 4 $\times$ 4 1-inch$^2$ S14161-6050HS SiPM array from Hamamatsu, 4 $\times$ 4 1-inch$^2$ J-60035 SiPM array from Onsemi, and 2 $\times$ 2 0.5-inch$^2$ S13370-6050CN SiPM array from Hamamatsu, were measured. Besides, a 3-inch R11065 PMT was used for comparison. More information about the four photosensors can be found in Tab.~\ref{table.1} and Fig.~\ref{fig:result_include1}. In the test, PMT has the sharpest SPE pulse shape and the effects of CT and AP are negligible compared with SiPM. While all the three kinds of SiPM show excellent SPE resolutions in a dark environment.

\section{Pulse construction using MC simulation} \label{sec:section5}

The PSD analysis is based on the Monte Carlo (MC) simulation and this section will introduce the details of the simulation. The nature of LAr scintillation is a distribution of photons in singlet states and triplet states with two decay times $\tau_s$ (the decay constant of the fast component) and $\tau_l$ (the decay constant of the slow component). Those parameters have been reported in multiple papers \cite{Deap2008, LAr_Luminescence, LArlight}, in which the results showed a distinct difference because they are derived from different liquid argon purities. The proportional coefficients of luminescence and decay times of different components in our simulation are based on the measurement result of a real single-phase LAr detector using PMT read-out.

\begin{figure}[htbp]
	\centering
	\includegraphics[width=7.5cm]{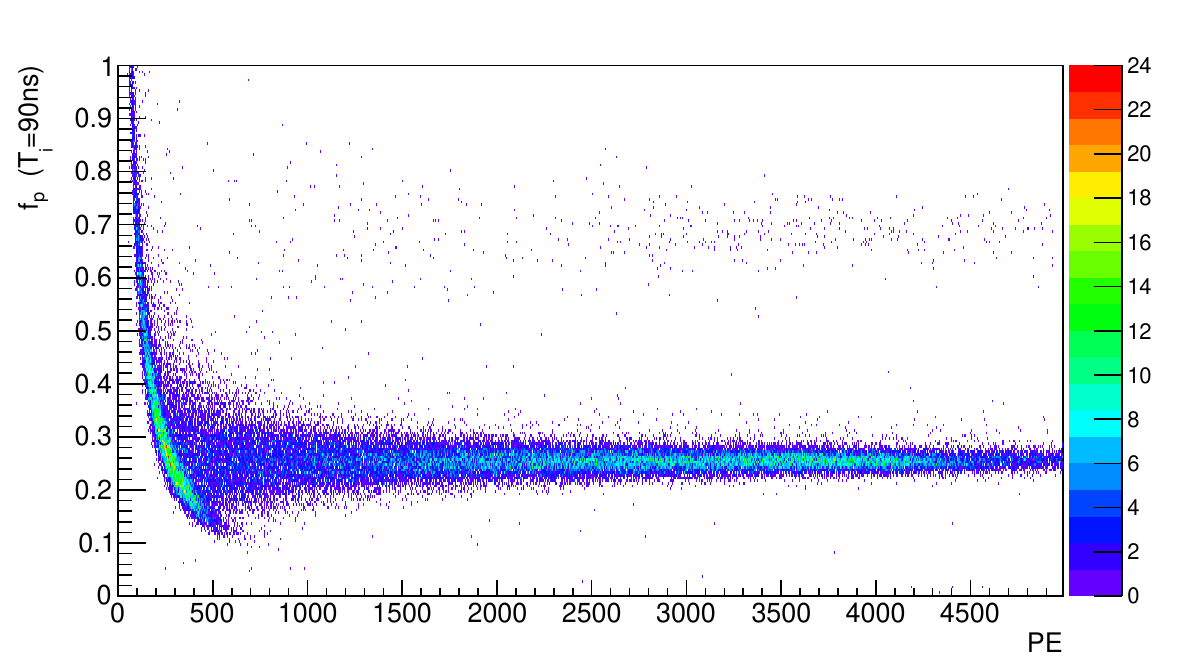}
	\qquad
	\includegraphics[width=7.5cm]{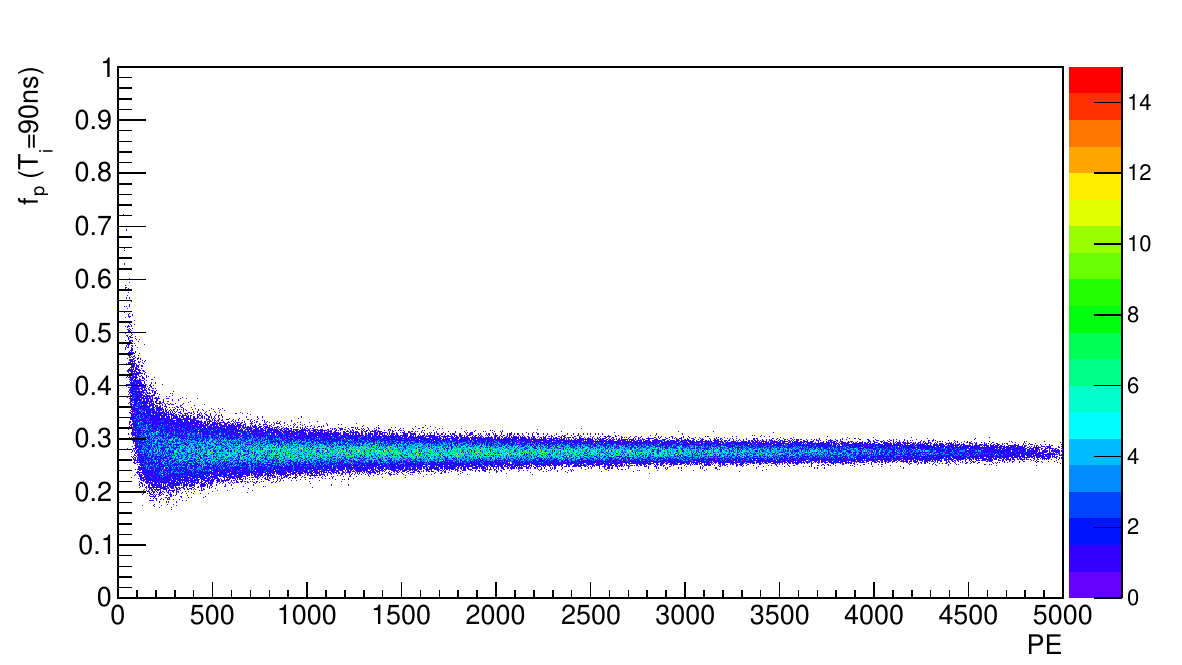}
	\caption{\label{PMT_F90} Top: Real data of f$_p$ distribution at T${_i}=90~ns$ in our single-phase LAr detector. Bottom: Simulated electron recoil distribution of f$_p$ at T${_i}=90~ns$.  } 	
\end{figure}

The body of the single-phase LAr detector is a cylinder with an inner diameter of 3~inches and a height of 3~inches. A 3-inch Hamamatsu R11065 PMT was placed at the bottom of the cylinder. The inner surface, except the PMT photocathode, was covered by Enhanced Specular Reflector film (ESR) to enhance the light collection. To improve the photon detection efficiency of PMT, the photocathode surface and all ESR surfaces were coated with 1,1,4,4-TetraPhenyl1-1,3-Butadiene (TPB) to convert the 128 nm photons to 420 nm. The upper plot in Fig.~\ref{PMT_F90} shows the f$_p$ spectrum with T${_i}$ = 90~ns from the region of 100~PE to 5000~PE. The obvious banded distribution around f$_p$(T${_i}$ = 90~ns) $\approx$ 0.3 represents electron recoil events, corresponding to $\gamma$ background scintillation in the LAr detector. The event stacking around 300 PE is caused by an $^{241}$Am radioactive source in the detector.

\begin{figure}[htb]
	\centering
	\includegraphics[width=7.5cm]{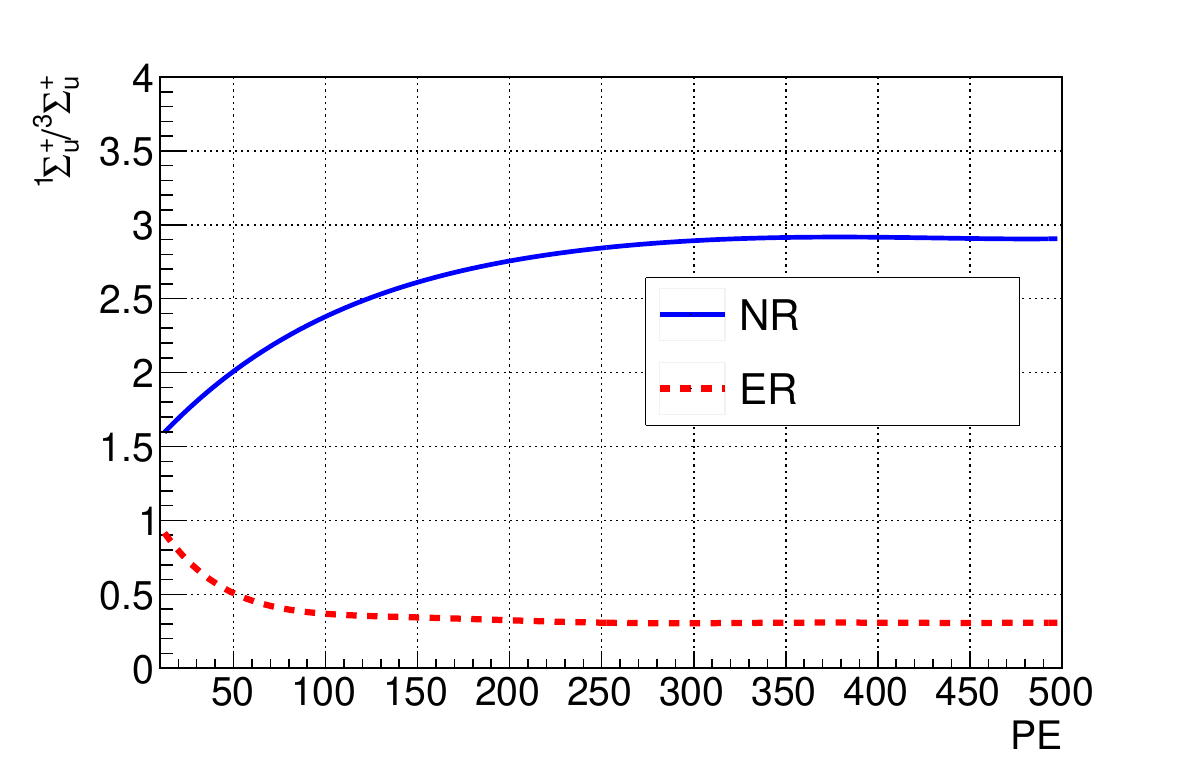}
	\caption{\label{FProportion} The function of the light yield ratio between the fast component ($^1 \Sigma_u^+$) and the slow component ($^3 \Sigma_u^+$) with the change of PE. The values of $\frac{^1 \Sigma_u^+}{^3 \Sigma_u^+}$ are from Ref.~\cite{Deap2008} but have been adjusted based on the measurement results of our single-phase LAr detector.} 	
\end{figure}

\begin{table}[htb]
\centering
\renewcommand{\arraystretch}{1.5}
\scriptsize
\begin{tabular}{ccc}
	
\hline
Parameter & Value & Remarks \\
	
\hline
$\tau_l$ & 1600~ns  & Decay time of $^3 \Sigma_u^+$\\
$\tau_s$ & 7~ns & Decay time of $^1 \Sigma_u^+$  \\
		
$f_{ER}$ & Red dotted curve in Fig.~\ref{FProportion}  & $\frac{^1 \Sigma_u^+}{^3 \Sigma_u^+}$ of electron recoils\\		
$f_{NR}$ & Blue solid curve in Fig.~\ref{FProportion} & $\frac{^1 \Sigma_u^+}{^3 \Sigma_u^+}$ of nuclear recoils\\
$\sigma_s^2$  &  0.55 ${\times}$ $N_{mean_s}$ & The diffusion factor of $^1 \Sigma_u^+$ \\
$\sigma_l^2$  &  0.97 ${\times}$ $N_{mean_l}$ & The diffusion factor of $^3 \Sigma_u^+$  \\
\hline 
\end{tabular}
\caption{\label{table.2} All parameters used in the simulation. The diffusion factors, $\sigma_s^2$  and $\sigma_l^2$ are the $\sigma$ of the Gaussian diffusion to smear the number of photo-electrons in reality. N$_{mean_s}$ and N$_{mean_l}$ are the calculated number of photo-electrons for the fast and slow components respectively. The diffusion factors determine the broadening of the striped region at the vertical dimension.
}
\end{table}


\begin{figure}[ht]
\centering
\includegraphics[width=13.5cm]{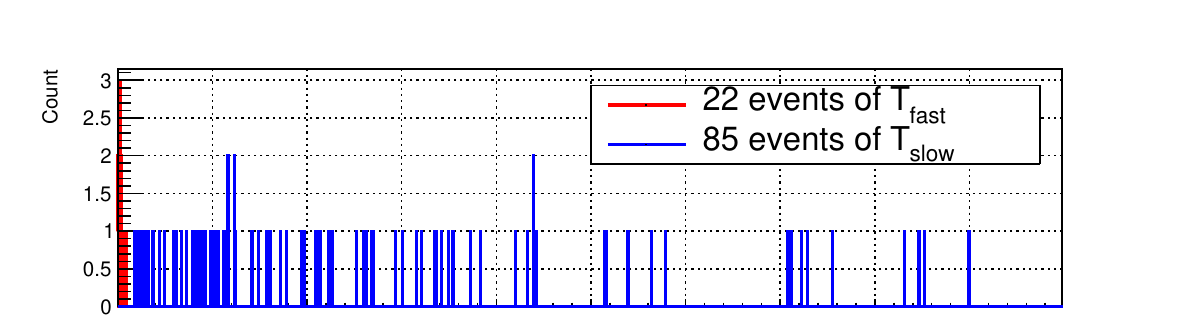}
\qquad
\includegraphics[width=13.5cm]{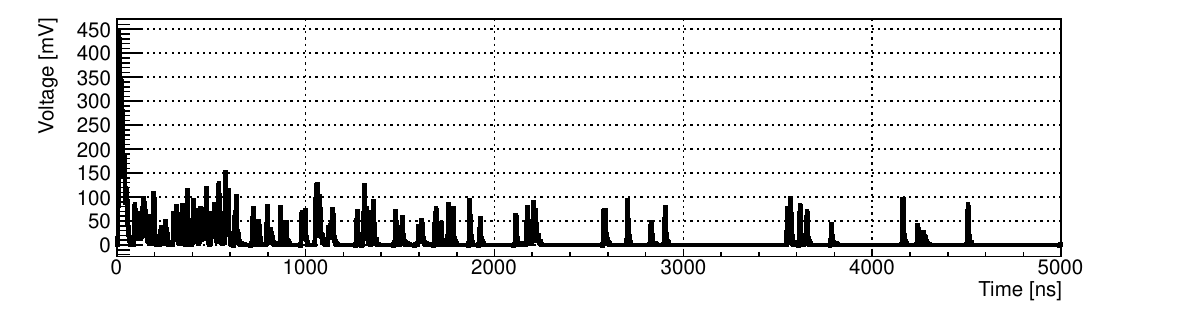}
\caption{\label{a_time_trace} Upper: A simulated hit-time pulse of a $
	\gamma$ event based on $^1 \Sigma_u^+ /^3 \Sigma_u^+ \approx$ 0.37 and $\tau_s\approx$ 7~ns, $\tau_l\approx$ 1600~ns, which has been converted to 22 photoelectrons of fast component (red) and 85 photoelectrons of slow component (blue) after a Gaussian diffusion. Lower: Using PMT SPE data measured in Sec.~\ref{sec:section3} to fill the hit-time pulse above creates a simulated PMT output pulse.} 	
\end{figure}

The SPE spectra of the Hamamatsu 8-inch R5912-20MOD PMT and three different SiPM products have been measured at liquid argon temperature, respectively. A program based on MC simulation has been developed to estimate the PSD capability of the LAr detector with PMTs or SiPM arrays as the photosensors using the SPE data.  The parameters used in the simulation are listed in Tab.~\ref{table.2}. The Lower plot in Fig.~\ref{PMT_F90} shows the simulated f$_p$ distribution. The difference in the low PE region between the two figures in Fig.~\ref{PMT_F90} is because the simulation did not consider the existence of $^{241}$Am radioactive source. The consistency of the f$_p$ distributions in the two plots of Fig.~\ref{PMT_F90} represents the reasonableness of the parameter values in the simulation.

\begin{figure}[htb]
\centering
\includegraphics[width=13.5cm]{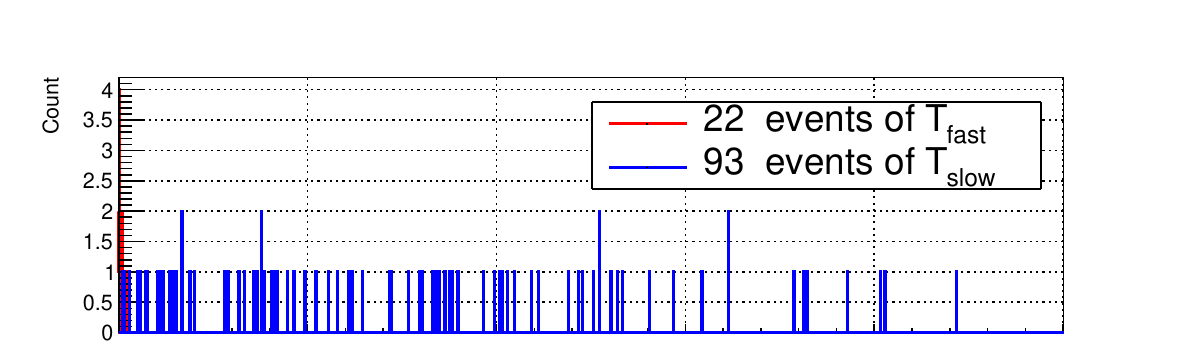}
\qquad
\includegraphics[width=13.5cm]{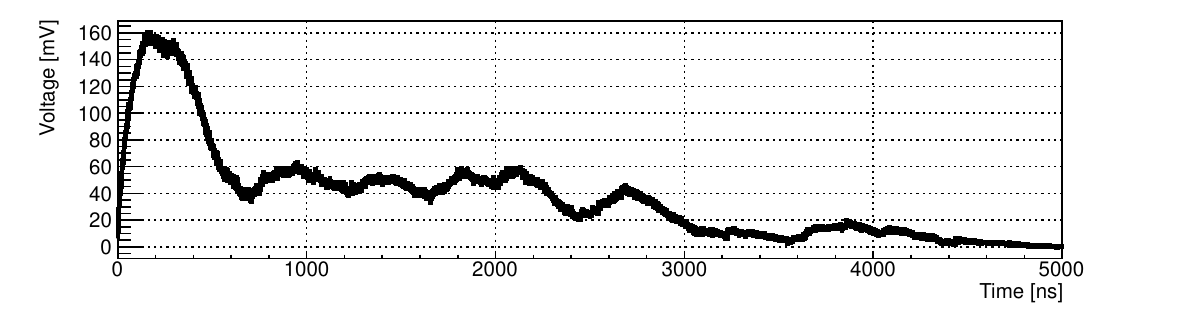}
\caption{\label{b_time_trace} An example of a simulated SiPM output pulse. Up: Hitting time pulse. Down: SiPM output pulse.} 	
\end{figure}

A hit-time pulse of the photosensors, which is similar to an output voltage pulse from the photosensors but only includes the hitting time information of each arrived photon, can be constructed with the light yield ratios and the decay times of the fast and slow components of LAr scintillation. For instance, for a 100 PE event, a hit-time pulse can be established according to the parameters introduced above. If this event is induced by $\gamma$, the light yield ratio for the fast and slow components is 0.37, then it will theoretically generate 27 fast component PEs and 73 slow component PEs. However, a Gaussian diffusion effect is inevitable to smear the number of PEs received by photosensors. $\sigma$ of Gaussian diffusion strongly depends on the detector performance specifically. The diffusion factors for the fast components ($\sigma_l$) and the slow components ($\sigma_s$) were defined as: $\sigma_l^2$ = 0.55${\times}$mean$_l$ and $\sigma_s^2$ = 0.97${\times}$mean$_s$ according to the PMT simulation result in Fig.~\ref{PMT_F90}. The upper plot in Fig.~$\ref{a_time_trace}$ shows a simulated hit-time pulse of a 100 PEs ER event and its 27 fast component PEs and 73 slow component PEs were smeared to 22 and 85. The PEs from the fast and slow components have been marked out by red and blue colors separately.

By filling the hit-time pulse with the PMT waveforms measured in the dark, a simulated voltage pulse of the LAr detector with PMT readout is generated, which is shown in the lower part of Fig.~$\ref{a_time_trace}$. Voltage pulses with SiPM readout can be constructed in the same way except using the SiPM arrays' waveforms measured in the dark. Fig.~$\ref{b_time_trace}$ shows an example of a simulated output voltage pulse. Since we uesd all the waveforms measured in the dark as the database, the effect of SiPMs' CT and AP probability are included in the simulation. The effect of SiPM's DCR is also included because a waveform is randomly selected from the same database to add on the baseline with the frequency shown in Tab.~\ref{table.1}.

\section{PSD capability calculation}\label{sec:section6}

\begin{figure}[htbp]
\centering
\includegraphics[width=9cm]{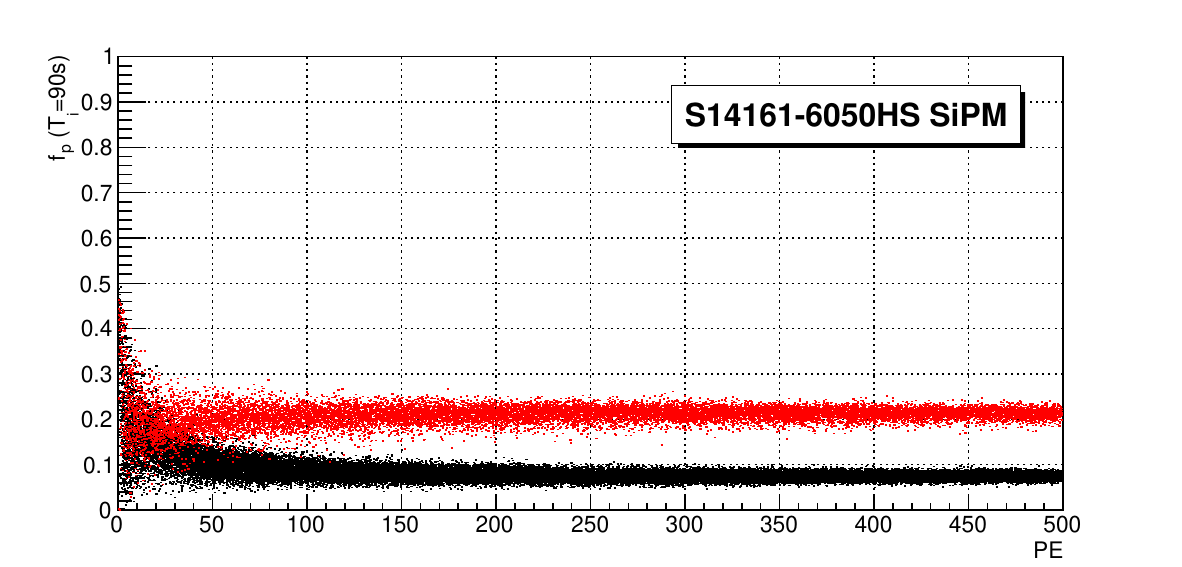}
\qquad
\includegraphics[width=9cm]{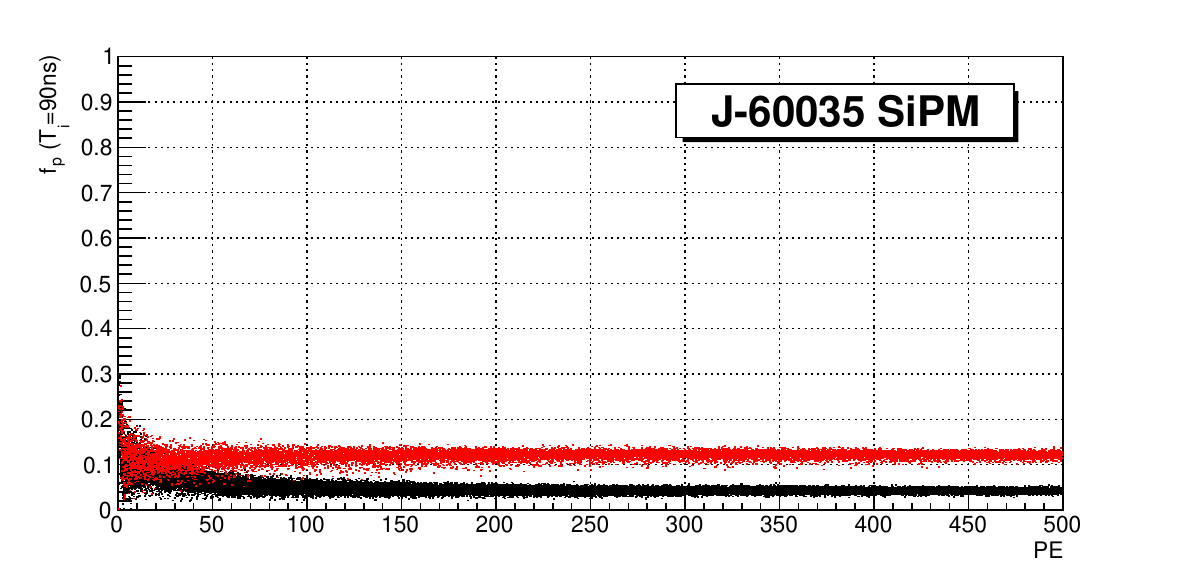}
\qquad
\includegraphics[width=9cm]{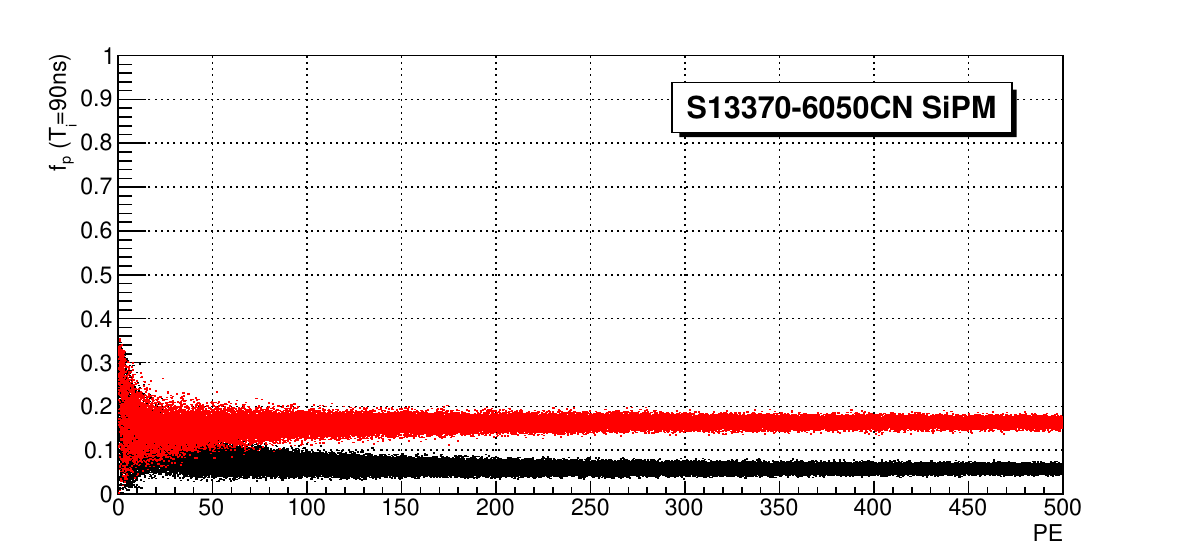}
\caption{\label{F90_Distribution} f$_p$ distributions change along with PE for LAr detector with three SiPMs as the photosensors. Three graphs include 100 thousand $\gamma$ events (black dots) and 50 thousand neutron events (red dots).}	
\end{figure}

The final objective of this simulation is to compare the PSD capability of LAr detectors when using PMT and three different SiPMs as photosensors based on the prompt fraction method. The MC simulation introduced at Sec.~\ref{sec:section5} successfully generated the LAr detector output pulses when using PMTs or SiPM arrays as photosensors with different PE numbers based on the waveforms collected in the dark. Fig.~\ref{F90_Distribution} shows the simulation results of f$_p$ distribution changes along with PE when using three different SiPM arrays, in which T$_i$ was set to 90~ns, T$_s$ and T$_e$ were set to 0~ns and 1~$\mu$s. Three simulations shared all the parameters except the waveform database. 

Three SiPM products have different DCR, CT and AP probabilities, SPE pulse shape, and SPE resolution. All of the characteristics would potentially affect the f$_p$ distributions theoretically. For example, J-60035 SiPM has the widest pulse shape, resulting in the smallest f$_p$ distribution, because the window of prompt 90~ns could only integral a very small part of the whole pulse. The phenomenon also indicates that T$_i$ = 90~ns is probably not the best choice for the prompt fraction method. The optimal value should depend on the specific SiPM type. 

\begin{figure}[htbp]  
		\centering          
	\subfigure[3-inch R11065 PMT, ERC = 2.47 $\pm$ 0.33 $\times$ 10$^{-4}$]{
	\begin{minipage}[t]{0.45\linewidth}
		\centering
		\includegraphics[width=1\linewidth]{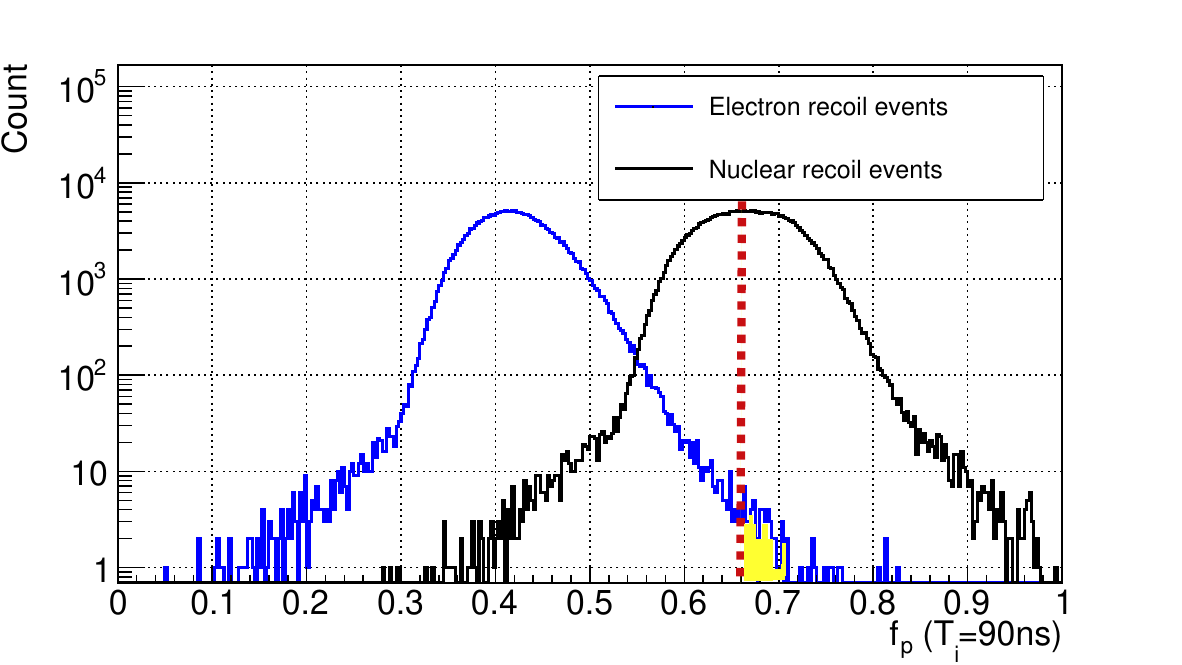}
	\end{minipage}
}
\subfigure[Sensl J-60035 SiPM, ERC = 9.48 $\pm$ 1.82 $\times$ 10$^{-5}$]{
	\begin{minipage}[t]{0.45\linewidth}
		\centering
		\includegraphics[width=1\linewidth]{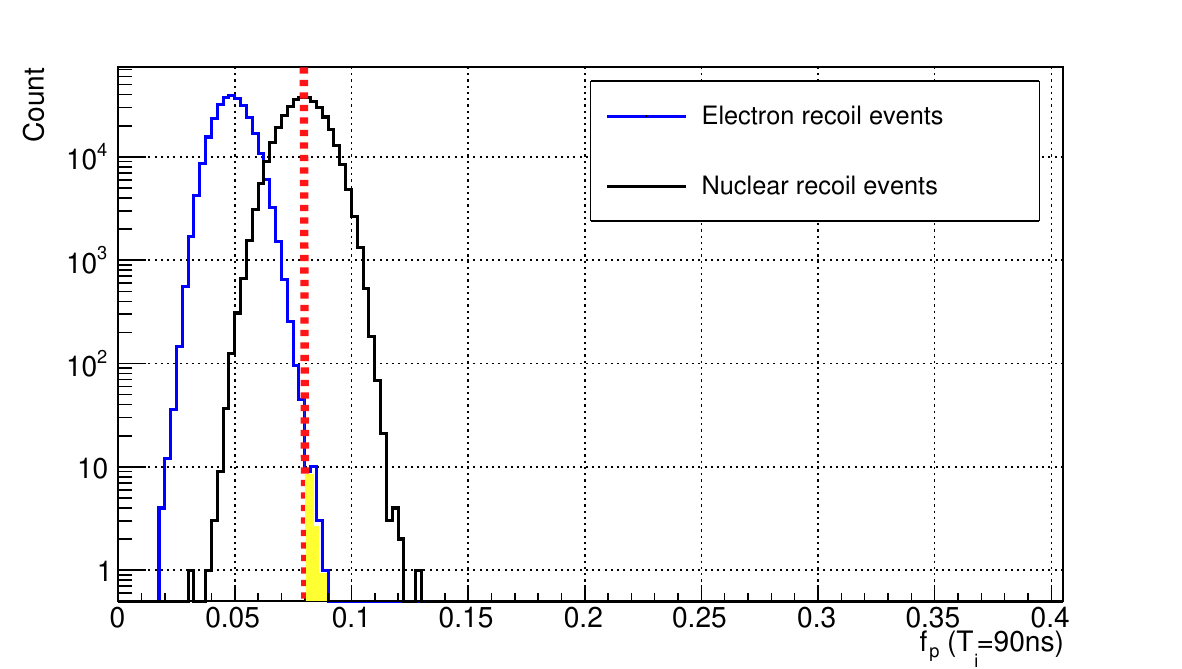}
	\end{minipage}
}
\subfigure[Hamamatsu S14161-6050HS SiPM, ERC = 1.21 $\pm$ 0.06 $\times$ 10$^{-3}$]{
	\begin{minipage}[t]{0.45\linewidth}
		\centering
		\includegraphics[width=1\linewidth]{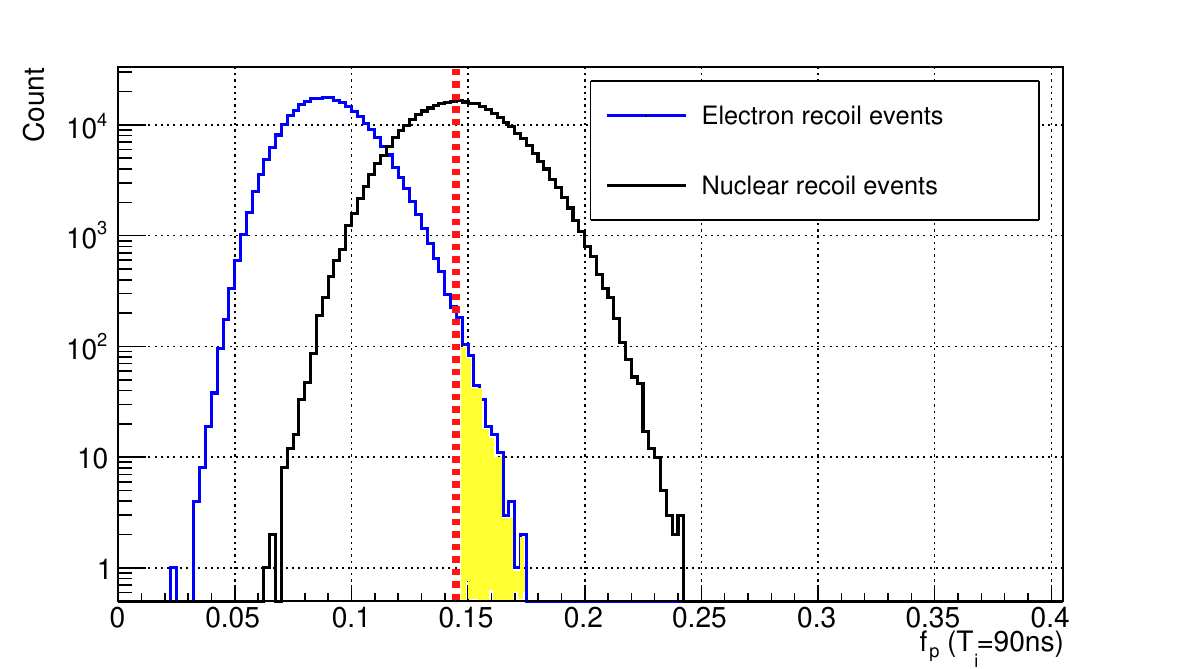}
	\end{minipage}
}
\subfigure[Hamamatsu S13370-6050CN SiPM, ERC = 7.57 $\pm$ 0.51 $\times$ 10$^{-4}$]{
	\begin{minipage}[t]{0.45\linewidth}
		\centering
		\includegraphics[width=1\linewidth]{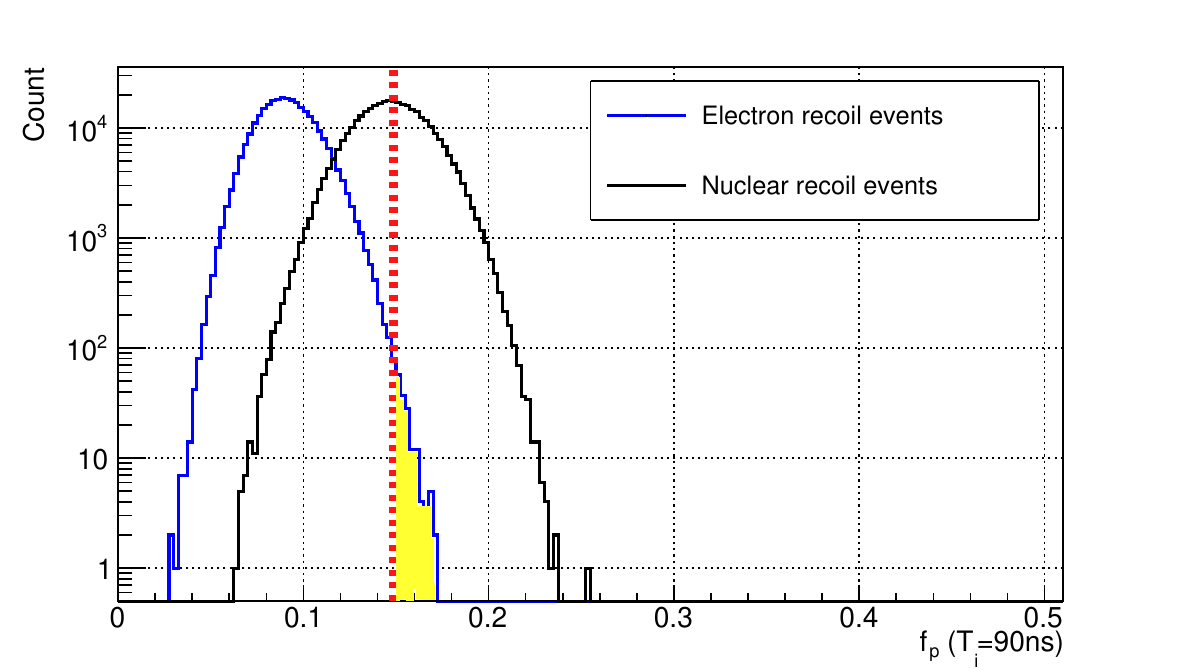}
	\end{minipage}
}
		\caption{Simulated results of f$_p$ distributions in the 28-32~PE energy range for LAr detector with PMTs or three SiPMs as the photosensors. ERC events are marked out with yellow areas. The ratio of the yellow area to the total area represents the ERC.}  
		\label{F90_Distribution_30PE}           
	\end{figure}
 
Another work in this paper is to optimize T$_i$ for each SiPM respectively. Here we use Electronic Recoil Contamination (ERC), defined by DEAP-3600~\cite{Deap2008}, to quantitatively estimate the PSD capability of different photosensors. It is the ratio of ER events of which f$_p$ are higher than the mean value of NR f$_p$ under a certain PE range.  DEAP-3600~\cite{Deap2008} achieved an ERC of around 2 $\times$ 10$^{-4}$ at 30 PE using PMTs. The simulation based on our R11065 PMT also gives an ERC result at 2.47 $\pm$ 0.33$\times$10$^{-4}$ at 28-32 PE energy range. 55 ERC events were picked out in 0.2 million ER events. The simulated f$_p$ distribution of ER and NR is shown in Fig.~\ref{F90_Distribution_30PE}(a). The ERC events are marked out as the yellow areas in Fig.~\ref{F90_Distribution_30PE}, and the ratio of the yellow area to the total area of ER represents the ERC value according to the definition.

\begin{figure}[htbp]
\centering
\includegraphics[width=10.5cm]{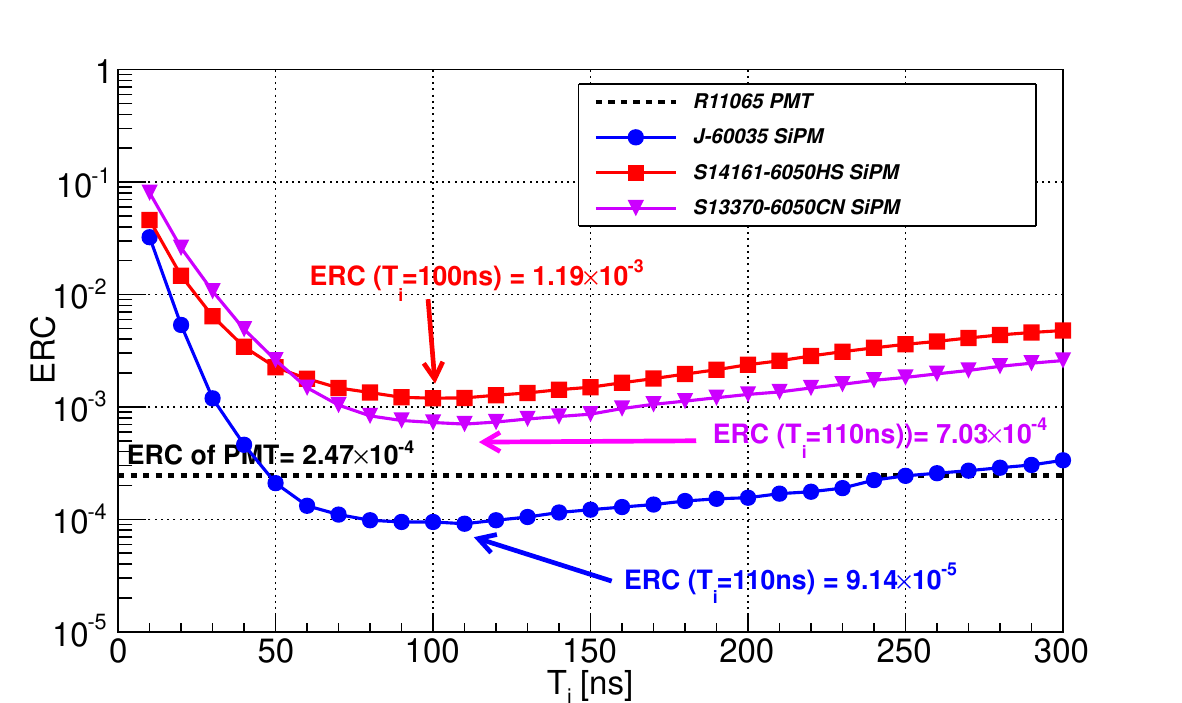}
\caption{\label{SiPMFp} ERC changes along with different integrating range T$_i$ for three SiPM products. The black dotted horizontal line corresponds to PMT ERC performance at $T_{i}$ = 90~ns as a comparison. The results show that different SiPM products should be setup with different $T_{i}$ to achieve the best ERC performance.}
\end{figure}

Fig.~\ref{SiPMFp} shows the ERC changes along with the integrating range $T_{i}$ in the 28-32 PE energy range. The black dotted horizontal line is the ERC of at $T_{i}$ = 90~ns. The other three curves with different colors represent the three kinds of SiPM arrays we tested. J-60035 SiPM array from Sensl shows the best ERC performance and it is the only SiPM product that has a better ERC performance than the PMT. The result shows that the much wider pulse like J-60035 SiPM's only causes the decrease of the distribution of f$_p$, while the PSD capability of a LAr detector does not get a negative effect from the slow rise and fall time of SPE pulses.

\section{Discussion}

As is shown in Fig.~\ref{SiPMFp}, the PSD capability of a LAr detector varies when different kinds of photosensors are used. The use of J-60035 SiPM results in a better PSD capability compared to the use of PMT as a photosensor, while using the other two kinds of SiPM deteriorates the PSD capability. The SiPM arrays offer a much better SPE energy resolution, which could benefit the PSD capability. However, the SiPM's drawbacks, namely the higher DCR and higher probabilities of CT and AP, could potentially decrease the PSD capability. 

According to the measurement results shown in Tab.~\ref{table.1} and Fig.~\ref{fig:result_include1} and the simulation results shown in Fig.~\ref{SiPMFp}, the S14161-6050HS SiPM has similar DCR and CT probability as the J-60035 SiPM, but the ERC is an order of magnitude worse. The main difference between the two SiPMs is the AP probability. APs appear after the primary pulse, so even APs produced by the fast component photons will contribute to the slow component while calculating the PSD capability with the prompt fraction method. Therefore, AP deteriorates the PSD capability of liquid argon detectors.

In order to verify the effect of DCR and CT on the PSD of the liquid argon detector, two additional simulations were done in the same way as described in Sec.~\ref{sec:section5}, except that once the waveforms of DCR were no longer randomly added to the baseline, and once only the waveforms of the SPE were taken into account while sampling. The ERC calculation results for these two cases are 8.12 $\times$ 10$^{-5}$ and 2.03 $\times$ 10$^{-5}$ with T$_i$ = 110~ns. It can be seen that both DCR and CT have negative effects on PSD capability. For the DCR, it is uniformly distributed throughout the time window. While calculating f$_p$, T$_i$ is much smaller than T$_e$, thus the contribution of DCR to the slow component is much greater than that to the fast component. For CT, it behaves as a multiphoton signal, so theoretically, it does not affect the energy distribution in the time window, however, for multi-PE instances, CT deteriorates the energy resolution of the detector~\cite{SLArD}, which will lead to a worse PSD capability of the detector.

\section{Conclusions}

SiPM is a new kind of photosensor that is expected to replace the traditional PMT for photon detection in the next generation of LAr detectors. The main objective of this paper is to compare the PSD capability of a LAr detector between using SiPM arrays as photosensors and using PMTs as photosensors. Three different SiPM arrays and one traditional PMT were used for comparison. The SPE performances of all samples were measured at liquid argon temperature in the dark.  The results show that a LAr detector could have the best PSD capability when using the J-60035 SiPM as the photosensor and the integral time for the fast component T$_i$ should increase slightly for SiPM readout LAr detectors. Our results also demonstrate that both AP and CT degrade the PSD capability of LAr detectors. However, it is observed that AP has a more significant impact compared to CT, primarily due to its influence on the energy distribution within the time window, while CT primarily affects the energy resolution of the detector.

\section{Acknowledgments}

This work is supported by the National Natural Science Foundation of China (Grant No.12275289, Grant No.11975257, and Grant No.12175247), the Youth Innovation Promotion Association of Chinese Academy of Sciences (Grant No.2023015), and the National Key R\&D Program of China (Grant No. 2016YFA0400304).



\begin{thebibliography}{00}
\bibitem{DarkSide-20k} DarkSide Collaboration, Darkside-20k: A 20 Tonne Two Phase LAr Tpc for Direct Dark Matter Detection at LNGS, European Physical Journal plus, 133(2018), pp.129.

\bibitem{Deap2008} W.H.~Lippincott, K.J.~Coakley, D.~Gastler, A.~Hime, E.~Kearns, D.N. McKinsey, J.A.~Nikkel, L.C.~Stonehill, Scintillation time dependence and pulse shape discrimination in liquid argon, Physical review. C,78(2008), p.035801

\bibitem{Gerda} GERDA Collaboration, Final results of GERDA on the Search for Neutrinoless Double-$\beta$ Decay,  Physical review letters, 125(2020), p.252502.

\bibitem{DUNE1} DUNE Collaboration, Long-baseline neutrino facility ({LBNF}) and deep underground neutrino experiment ({DUNE}) conceptual design report volume 1: The LBNF and DUNE projects, White Rose Research Repository (2016).

\bibitem{PSDpaper1} B.G~Boulay and A.~Hime, Technique for direct detection of weakly interacting massive particles using scintillation time discrimination in liquid argon, Astroparticle physics, 25(2006), pp.179-182.

\bibitem{PSDpaper2} DEAP Collaboration, Measurement of the scintillation time spectra and pulse-shape discrimination of low-energy $\beta$ and nuclear recoils in liquid argon with DEAP-1, Astroparticle physics, 85(2006), pp.1-23.

\bibitem{PSDpaper3} DarkSide Collabration, First results from the {D}ark{S}ide-50 dark matter experiment at Laboratori Nazionali del Gran Sasso, PHYSICS LETTERS B, 743(2015), pp.456-466.

\bibitem{SiPM} D.~Renker, Geiger-mode avalanche photodiodes, history, properties and problems,  Nucl.~Instrum.~Meth.~A, 567(2006), pp.48-56.

\bibitem{nEXO} I.~Ostrovskiy, F.~Retiere, D.~Auty, J.~Dalmasson, T.~Didberidze, R.~DeVoe, G.~Gratta, L.~Huth, L.~James, L.~Lupin-Jimenez, N.~Ohmart, A.~Piepke, Characterization of silicon photomultipliers for nEXO, IEEE transactions on nuclear science, 62(2015), pp.1825-1836.

\bibitem{PreviousWork1} N.~Canci, C.~Cattadori, M.~D'Incecco, B.~Lehnert, A.A.~Machado, S.~Riboldi, D.~Sablone, E.~Segretoc, C.~Vignolic, Liquid argon scintillation read-out with silicon devices, Journal of instrumentation, 8(2013), p.C10007.

\bibitem{PreviousWork2} M.~Boulay, B.~Cai, M.~Chen,V.~Golovko, P.~Harvey, R.~Mathew, J.~Lidgard, A.~McDonald, P.~Pasuthip, T.~Pollman, P.~Skensved, K.~Graham, A.~Hallin, D.~Mckinsey, W.~Lippincott, J.~Nikkel, C.~Jillings, F.~Duncan, B.~Cleveland, I.~Lawson, Cryogenic characterization of FBK RGB-HD SiPMs, Journal of instrumentation, 12(2017), p.P09030.

\bibitem{PreviousWork3} E.~Segreto, A.A~ Machado, L.~Paulucci, F.~Marinho, D.~Galante, S.~Guedes, A.~Fauth, V.~Teixeira, B.~Gelli, M.R.~ Guzzo, W.~Araujo, C.~Ambrósio, M.~Bissiano, A.L.~Lixandrão Filho, Liquid argon test of the ARAPUCA device, Journal of instrumentation, 13(2018), p.P08021.

\bibitem{LAr_Luminescence} A.~HITACHI, T.~TAKAHASHI, N.~FUNAYAMA, K.~MASUDA, J.~KIKUCHI, and T.~DOKE, Effect of ionization density on the time dependence of luminescence from liquid argon and xenon, Physical review. B, 27(1983), pp.5279-5285.

\bibitem{PreviousWorks4}T.A.~Wang, C.~Guo, X.H.~Liang, L.~Wang, M.Y.~Guan, C.G.~Yang, J.C.~Liu, F.Y.~Lin, Characterization of two SiPM arrays from Hamamatsu and Onsemi for liquid argon detector, Nucl.~Instrum.~Meth.~A, 1053 (2023) 168359.

\bibitem{VUV4_SiPM} L.~Wang, M.Y.~Guan, H.J.~Qin, C.~Guo, X.L.~Sun, C.G.~Yang, Q.~Zhao, J.C.~Liu,P.~Zhang, Y.P.~Zhang, W.X.~Xiong, Y.T.~Wei, Y.Y.~Gan, J.J.~Li, Characterization of VUV4 SiPM for liquid argon detector,  Journal of instrumentation, 16(2021), p.P07021.

\bibitem{XENON1T} E.~Aprile, J.~Aalbers, A.P~Colijn, M.P.~Decowski, et~al., Xenon1t dark matter data analysis: Signal and background models and  statistical inference, Physical review. D, 99(2019), p.1.

\bibitem{HamaDS} Hamamatsu R5912-20MOD PMT datasheet, https://www.hamamatsu.com.

\bibitem{SenslDS} Sensl J-Series SiPM datasheet, https://www.onsemi.com.

\bibitem{Amp} LMH6629 datasheet, https://www.ti.com.

\bibitem{LArlight} E.~Segreto, Properties of liquid argon scintillation light emission, Physical review. D, 103(2021), p.043001.

\bibitem{SLArD} L.~Wang, et al., Developing a single-phase liquid argon detector with SiPM readout, Eur. Phys. J. Plus (2023) 138:629.

\end{thebibliography}



\end{document}